# On the Performance of Multihop-Intervehicular Communications Systems over n*Rayleigh Fading Channels

Yahia Alghorani, *Student Member, IEEE*, Georges Kaddoum, *Member, IEEE*, Sami Muhaidat, *Senior Member, IEEE*, Samuel Pierre, *Senior Member, IEEE,* and Naofal Al-Dhahir, *Fellow, IEEE*

*Abstract*—We investigate the performance of multihop-intervehicular communication systems with regenerative and nonregenerative relaying. We consider the so-called "*n*\*Rayleigh distribution" as an adequate multipath fading channel model for vehicle-to-vehicle communication scenarios. We derive a novel approximation for the outage probability of maximum ratio combining (MRC) diversity reception. In addition, we analyze the amount of fading and optimize the power allocation for the investigated scenario. Numerical results show that regenerative systems are more efficient than nonregenerative systems when the cascading order (*n*) is small; however, for large *n*, our results demonstrate that the performance of both relaying techniques is rather similar.

*Index Terms*—Multihop, *n*\*Rayleigh fading, MRC, diversity, intervehicular communications.

## I. INTRODUCTION

MULTIHOP transmission is advantageous when the distance between the source and destination is large, as it can be used to extend the coverage without using large transmit power [1, 2].

In intervehicular communication (IVC) systems, both the transmitter and receiver are in motion and typically use the same antenna height, resulting in two or more independent Rayleigh fading processes, generated by independent groups of scatterers around the two mobile terminals [3]. Such kind of keyhole propagation scenarios is possible when two rings of scatterers separated by a large distance and all propagation paths travel through the same narrow pipe called "*n*\*Rayleigh fading channels"[4]. A special case of this fading model was studied in [5] when double Rayleigh fading is considered (i.e., the cascading order $n = 2$). The multiple-input multiple-output (MIMO) case was also discussed in [6]. In [7], experimental results in different vehicular communication environments have shown that in vehicular networks, several small-scale fading processes are multiplied together, leading to a worse-than Rayleigh fading. Several studies in the literature have also analyzed the V2V channel characteristics (i.e., path loss, delay spread, Doppler spread, level crossing rate (LCR) and average fade duration (AFD)) using the so-called "double-ring geometric model" to simulate the mobile-to-mobile local scattering environment [8]. Although the geometric models can be used to accurately model the V2V channel characteristics in a wide variety of environments, unfortunately, they are complex and require numerous parameter selections for the specific environment of interest [9].

Taking advantage of cooperative diversity systems across generalized fading channels, there have been some recent studies that have investigated dual hop IVC systems with relay selection strategy, e.g., [10, 11]. However, all results have been reported over double-Rayleigh/Nakagami fading channels. To the best of the authors' knowledge, multihop systems with maximum ratio combining (MRC) schemes over *n*\*Rayleigh fading channels (i.e, $n \geq 2$) have not been analyzed yet. Therefore, it is the aim of this work to fill this research gap and investigate the performance of multihop-IVC systems with MRC diversity reception under *n*\*Rayleigh distribution.

## II. SYSTEM MODEL

We consider a $N$-hop intervehicular communications system where the source ($s$) transmits information to its destination ($d$) through intermediate nodes $r_i$ ($i = 1, 2, ...., N - 1$), acting as regenerative or nonregenerative relays. All underlying $N$ channels/hops are modeled as a product of $n$ independent circularly-symmetric complex Gaussian random variables, each of which can be defined as $h_i \triangleq \prod_{j=1}^{n} h_{i,j}$ with zero mean and channel variance $\lambda_i$. Thus, $|h_i|$ follows the *n*\*Rayleigh distribution. We assume that the received signal undergoes slow fading (i.e., the symbol period of the received signal is smaller than the coherence time of the channel), which can be justified for rush-hour traffic. We further assume that the additive white Gaussian noise (AWGN) random processes at all relays and the destination node have zero mean and variance ($N_o$). In this case, the instantaneous signal-to-noise ratio (SNR) of the *i*th hop is given by $\gamma_i = |h_i|^2 P/N_o$, where $P$ is the transmitted signal power. An accurate approximation for the probability density function (PDF) of the instantaneous SNR $\gamma_i$ is given by [12] with the help of [13, eq.(2.3)]

$$f_{\gamma_i}(\gamma) \approx \frac{b_i^{m_i}\gamma^{a_i-1}}{n_i\Gamma(m_i)}\exp\left(-b_i\gamma^{\frac{1}{n_i}}\right), \quad \gamma \geq 0 \quad . \quad (1)$$

where $\Gamma(.)$ represents the Gamma function, defined in [14, eq.(8.310.1)], $a_i = \frac{m_i}{n_i}$, and $b_i = 2m_i/\Omega_i\bar{\gamma}_i^{1/n_i}$ where $\bar{\gamma}_i = \lambda_i P/N_o$ is the average SNR of the *i*th hop, $\lambda_i = \mathbf{E}(|h_i|^2)$ with $\mathbf{E}(.)$ denoting expectation. The fading severity parameters of the *i*th hop $(m_i, \Omega_i)$ are positive real numbers given by [12]

$$m_i = 0.6102n_i + 0.4263, \quad \Omega_i = 0.8808n_i^{-0.9661} + 1.12$$

It is important to note that the novel approximation for the PDF in (1) was examined in [12], by comparing it to the exact PDF derived in [15, eq.(8)] and showing that the new approximation has high accuracy in most cases considered. Furthermore, the approximate PDF is easy to calculate and to manipulate compared to the exact PDF.

Assume that each node in the considered multihop scheme is equipped with $L$ diversity branches, then the total SNR at the output of the MRC combiner is given by $\gamma_{ti} = \sum_{l=1}^{L} \gamma_{i,l}$. Normalizing (1) by the PDF of the SNR at the MRC output with Nakagami fading [16, eq.(7)] and replacing the factor $2m_i/\Omega_i$ by $2Lm_i/\Omega_i$, and $\bar{\gamma}_i$ by $L\bar{\gamma}_i$, the approximate PDF of the combined SNR $\gamma_{ti}$ for the *i*th hop over independent and identically distributed (i.i.d) *n*\*Rayleigh fading random variables, can be derived as

$$f_{\gamma_{ti}}(\gamma_t) \approx \frac{\beta_i^{Lm_i}\gamma_t^{\alpha_i-1}}{n_i\Gamma(Lm_i)}\exp\left(-\beta_i\gamma_t^{\frac{1}{n_i}}\right). \quad (2)$$

where $\alpha_i = Lm_i/n_i$ and $\beta_i = 2 Lm_i/\Omega_i(L\bar{\gamma}_i)^{1/n_i}$. However, eq.(2) is novel and has not been derived yet for the MRC scheme with

Y.Alghorani and S.Pierre are with the Mobile Computing and Networking Research Laboratory (LARIM), Department of Computer Engineering, Ecole Polytechnique de Montreal, Station Centre-ville, Montreal, Quebec, H3C 3A7, Canada (e-mail: yahia.alghorani@polymtl.ca; samuel.pierre}@polymtl.ca).

G.Kaddoum is with the Electrical Engineering Department, École de Technologie Supérieure, Montréal, Canada, H3C1K3, (email: Georges.Kaddoum@etsmtl.ca).

S. Muhaidat is with the Electrical and Computer Engineering Department, University of Western Ontario, London, Ontario, N6A 5B9 , Canada, (e-mail: sami.muhaidat@uwo.ca).

N. Al-Dhahir is with the Department of Electrical Engineering, The University of Texas at Dallas, Richardson, TX 75083 USA (e-mail: aldhahir@utdallas.edu).

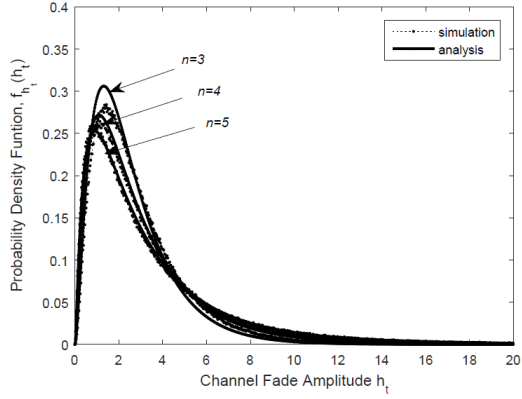

Fig. 1. Comparison between analytical results and Monte-Carlo simulation for the PDF formulated by (3) ($n = 3; m = 2.256, \Omega = 1.424$), ($n = 4; m = 2.87, \Omega = 1.351$), ($n = 5; m = 3.477, \Omega = 1.306$), $\sigma^2 = 1$, $L = 2$ and $10^6$ iterations.

$n$*Rayleigh fading. In addition, by introducing a change of variables in the expression for the PDF $f_{\gamma_{ti}}(\gamma_t)$ of $\gamma_{ti}$, $f_{\gamma_{ti}}(\gamma_t) = f_{h_t}(\sqrt{2^n \sigma_i^2 \gamma_t / \bar{\gamma}_i})/2\sqrt{\gamma_t \bar{\gamma}_i/2^n \sigma_i^2}$, the channel fading amplitude $h_t$ is distributed according to

$$f_{h_{ti}}(h_t) \approx \frac{2\left(\frac{Lm_i}{\Omega_i}\right)^{Lm_i} h_t^{2\alpha_i - 1}}{n_i \Gamma(Lm_i)(L\sigma_i^2)^{\alpha_i}} \exp\left(-\frac{Lm_i}{\Omega_i}\left(\frac{h_t}{L\sigma_i}\right)^{\frac{2}{n_i}}\right). \quad (3)$$

Where $\sigma_i$ is the standard deviation of the original complex Gaussian signal prior to envelop detection at each hop, the parameter $\sigma_i^2 = \prod_{j=1}^{n} \sigma_{i,j}^2$, and $2^n \sigma_i^2 = \lambda_i$. As a double check, the PDF in (3) is validated by Monte-Carlo simulation, as observed in Fig. 1, the approximate PDF has high accuracy as the cascading order $n$ increases. The larger the value of $n$ is, the higher accuracy will be [12].

For nonregenerative relaying that is employed in analog systems, the relay amplifies the incoming signal and forwards it to the next relay without decoding. In this case, the end-to-end SNR, $\gamma_{eq}$, at the destination can be upper-bounded by [1]

$$\gamma_{eq} = \left(\sum_{i=1}^{N} \frac{1}{\gamma_{ti}}\right)^{-1}. \quad (4)$$

Since (4) is related to the harmonic mean of individual links SNRs, $\gamma_{ti}$, we can use the following inequality proposed by [2]

$$\gamma_{eq} \leq \frac{1}{N} \prod_{i=1}^{N} \gamma_{ti}^{\frac{1}{N}}. \quad (5)$$

From (5), the $k$th moment of $\gamma_{eq}$ over identical fading severity parameters can be evaluated with the help of [14, eq.(3.326.2)], as

$$\mathbf{E}(\gamma_{eq}^k) \leq \frac{N^{-k}}{\prod_{i=1}^{N} \beta_i^{\frac{nk}{N}}} \left[\frac{\Gamma\left(\frac{nk}{N} + Lm\right)}{\Gamma(Lm)}\right]^N. \quad (6)$$

where $\beta_i = 2Lm/\Omega(L\bar{\gamma}_i)^{1/n}$. By using the *inverse Mellin transform* of $\mathbf{E}(\gamma_{eq}^k)$, defined by the contour integral $f_X(x) = \int_L \frac{1}{j2\pi} \mathbf{E}(X^k) x^{-(k+1)} dk$ [14, Sec.17.41] and with the help of the Meijer's G-function identity in [17, eq.(07.34.02.0001.01)], the upper-bound PDF and the cumulative density function (CDF) of the end-to-end SNR can be expressed as

$$f_{\gamma_{eq}}(\gamma_t) \leq \frac{\gamma_t^{-1} G_{0,N}^{N,0}\left((N\gamma_t)^{\frac{N}{n}} \prod_{i=1}^{N} \beta_i \,\Big|\, {}_{Lm,\ldots,Lm}^{-}\right)}{\Gamma^N(Lm)} \quad (7)$$

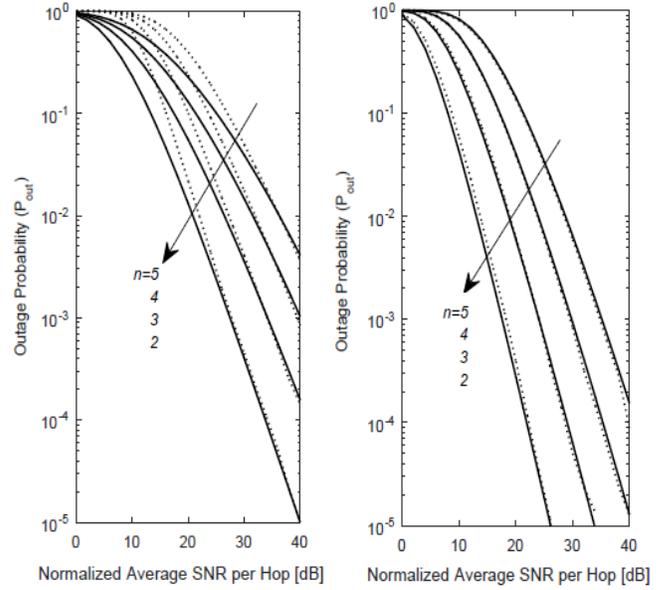

Fig. 2. Comparison between analytical results and Monte-Carlo simulation for the outage probability lower-bounds of regenerative and nonregenerative systems with MRC diversity reception over $n$*Rayleigh fading channels (Solid lines: analysis, dotted lines: simulation). Left: the left tails of the outage probability of nonregenerative systems ($N = 4, L = 2$). Right: the right tails of the outage probability of regenerative systems ($N = 6, L = 3$).

and

$$F_{\gamma_{eq}}(\gamma_t) \leq \frac{G_{1,N+1}^{N,1}\left((N\gamma_t)^{\frac{N}{n}} \prod_{i=1}^{N} \beta_i \,\Big|\, {}_{Lm,\ldots,Lm,0}^{1}\right)}{\Gamma^N(Lm)}. \quad (8)$$

respectively, where $G_{p,q}^{m,n}(.)$ is the Meijer's G-function defined in [14, eq.(9.301)]. Note that the inverse Mellin transform approach has the advantage of simplicity in the derivation of both (7) and (8) rather than using the moment generating function (MGF) approach as in [2].

*Special cases:* For $N = 1$ with the help of [17, eq.( 07.34.03.0228.0 1)], (7) simplifies to (2). In addition, for $N = 2$ with the help of [17, eq.(07.34.03.0605.01)], (7) reduces to the following case of dual-hop transmission.

$$f_{\gamma_{eq}}(\gamma_t) \leq \frac{2\gamma_t^{-1}\left((2\gamma_t)^{\frac{2}{n}} \prod_{i=1}^{2} \beta_i\right)^{Lm} K_0\left(2(2\gamma_t)^{\frac{2}{n}} \prod_{i=1}^{2} \beta_i\right)}{\Gamma^2(Lm)}. \quad (9)$$

where $K_0(\cdot)$ is the zeroth-order modified Bessel function of the second kind defined in [14, eq.(9.6.21)].

For regenerative relaying that is employed in digital systems, the relay decodes the received signal and then forwards it to the next hop [1]. In this case, the underlying scheme takes a decision per hop and the equivalent SNR is $\gamma_{eq} = \min\{\gamma_{t1}, \ldots, \gamma_{tN}\}$, which leads to derive the approximate CDF of $\gamma_{eq}$ as follows

$$F_{\gamma_{eq}}(\gamma_t) = \Pr\left(\min_{i \in N}\{\gamma_{ti}\} \leq \gamma_t\right)$$
$$= 1 - \Pr(\gamma_{t1} > \gamma_t, \gamma_{t2} > \gamma_t, \ldots, \gamma_{tN} > \gamma_t) \quad (10)$$

Using the fact defined in [14, eq.(3.381.1)], (10) can be expressed as

$$F_{\gamma_{eq}}(\gamma_t) \approx 1 - \prod_{i=1}^{N} \frac{\Gamma\left(Lm_i, \beta_i \gamma_t^{\frac{1}{n}}\right)}{\Gamma(Lm_i)}. \quad (11)$$

where $\Gamma(.,.)$ represents the upper incomplete gamma function, defined by $\Gamma(\alpha, x) = \int_x^\infty e^{-t} t^{\alpha-1} dt$ [14].

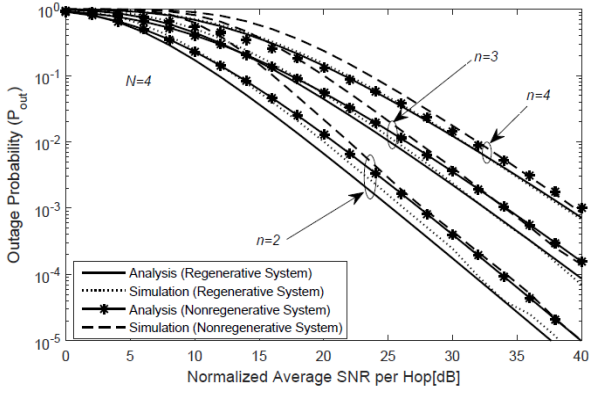

**Fig. 3.** End-to-end outage probability of regenerative and nonregenerative systems with MRC diversity reception over $n*$Rayleigh fading channels ($L = 2$).

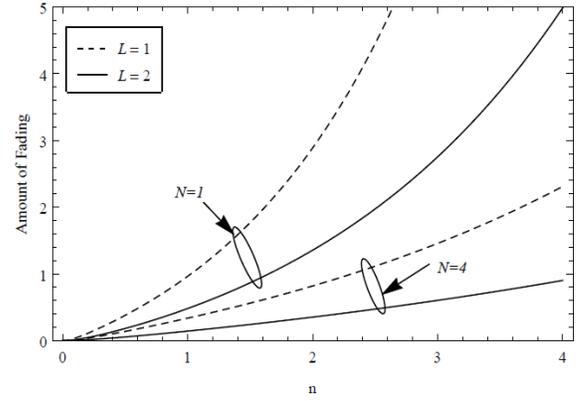

**Fig. 4.** Amount of fading of direct and multihop transmission systems with diversity reception over n*Rayleigh fading channels.

## III. PERFORMANCE ANALYSIS

### A. Outage Probability

The outage probability of a communication channel can be defined as the probability that the end-to-end SNR $\gamma_{eq}$ falls below a certain threshold $\gamma_{th}$, namely

$$P_{out} = \Pr(\gamma_{eq} \leq \gamma_{th}) = F_{\gamma_{eq}}(\gamma_{th}) . \quad (12)$$

Using (8) and (11), lower-bounds for the outage probability can be derived for nonregenerative and regenerative systems respectively. Furthermore, using (8) and the asymptotic expression for the Meijer's G-function presented in [17, eq. (07.34.06.0006.01)], we can derive an approximate closed-form expression for the outage probability of nonregenerative systems in the high-SNR regime (i.e., when $\bar{\gamma}_i \to \infty$). On the other hand, assuming that the instantaneous SNRs for all hops are i.i.d random variables (i.e., $\bar{\gamma}_i = \bar{\gamma}$) in (11), we can derive a simple lower-bound expression for the outage probability of regenerative systems in the high-SNR regime using the fact that $\Gamma(\alpha, x) = \Gamma(\alpha) - \sum_{n=0}^{\infty} \frac{(-1)^n x^{\alpha+n}}{n!(\alpha+n)}$ [14], which yields $\Gamma(\alpha, x) = \Gamma(\alpha) - x^\alpha/\alpha$ when $x \to 0$, as

$$P_{out} \approx \frac{N\left(\beta\gamma_t^{\frac{1}{n}}\right)^{Lm}}{Lm\Gamma(Lm)} . \quad (13)$$

From (13), it is noted that the end-to-end outage probability increases as a number of hops increases and it significantly decreases when diversity combining schemes are employed. Fig. 2 shows the outage probability achieved by both schemes, which are validated by Monte-Carlo simulation. As clearly observed from Fig. 2, lower-bounds for the outage probability-based (8) and (11) converge to simulation results in the high SNR regime. Additionally, the larger the value of $n$ is, the tighter the bounds are obtained.

Fig.3 compares the outage probability of a regenerative system with that of a nonregenerative system over $n*$Rayleigh fading channels. As we can see in Fig.3, the approximation error of the bound is obviously noticed for $n = 4$ with low-to-medium SNR range (i.e, SNR$\leq 20$ dB) and the accuracy of the bound gradually improves in the high-SNR regime. Furthermore, the regeneration improves the outage performance, and the performance difference between two systems starts decreasing gradually as the cascading order $n$ increases. This means that the decoding error probability for regenerative systems increases as the fading severity parameter $n$ increases. In this case, nonregenerative systems with diversity reception enhance the performance without increasing the complexity of systems design.

### B. Amount of Fading

The end-to-end amount of fading (AF) is defined as the ratio of variance to the square average SNR [13].

$$AF = \frac{\text{var}(\gamma_{eq}^2)}{\left(\mathbf{E}[\gamma_{eq}]\right)^2} .$$

where $\text{var}(.)$ denotes variance. Using (6), the following AF lower-bound can be derived

$$AF \approx \left[\frac{\Gamma(Lm)\Gamma\left(\frac{2n}{N} + Lm\right)}{\Gamma^2\left(\frac{n}{N} + Lm\right)}\right]^N - 1 . \quad (14)$$

*Special case:* For $N = 1$ and $L = 1$, AF can be simplified by

$$AF \approx \frac{(m)_{2n}}{(m)_n^2} - 1 . \quad (15)$$

where $(x)_n = \Gamma(x+n)/\Gamma(x)$. From (15), it is noted that for the classical Rayleigh distribution ($n = 1$), AF $\approx 1$.

Fig. 4 shows the effect of diversity on the performance of both direct and multihop transmission systems. It is interesting to note that: 1) using relays, the amount of fading is reduced compared to the direct transmission, 2) the overall amount of fading is reduced substantially when diversity reception is employed, and 3) the performance difference between diversity combining systems and no-diversity reception becomes larger as $n$ increases. These results provide new insight into the trade-off between performance and complexity diversity reception over $n*$Rayleigh fading channels and assist in the design of such receivers.

### C. Power Optimization

From a practical standpoint, tracking the signal/interference level during the real-time updates of power allocation (PA) among vehicles requires high hardware complexity. This motivates us to investigate optimized power allocation when perfect statistical channel state information (CSI) is available at the source and relay nodes.

To simplify the analysis, we only evaluate the power allocation mode for regenerative systems. In this case, we redefine the instantaneous SNR as $\bar{\gamma}_i = P_i\lambda_i$, where $P_i \triangleq E_{si}/N_o$ is the transmitted signal power per hop with $E_{si}$ denoting the transmitted signal energy. Next, we can optimize the power allocation to minimize the outage probability in (12) under a total power constraint ($\sum_{i=1}^{N} P_i \leq P_T$) with the knowledge of channel statistics $\lambda_i$. Consequently, the optimization problem is formulated as follows

$$\min_{P_i} \left(1 - \prod_{i=1}^{N} \frac{\Gamma\left(Lm_i, \beta_i\gamma_{th}^{\frac{1}{n_i}}\right)}{\Gamma(Lm_i)}\right)$$

subject to $\sum_{i=1}^{N} P_i \leq P_T$ and $P_i \geq 0$ \quad (16)

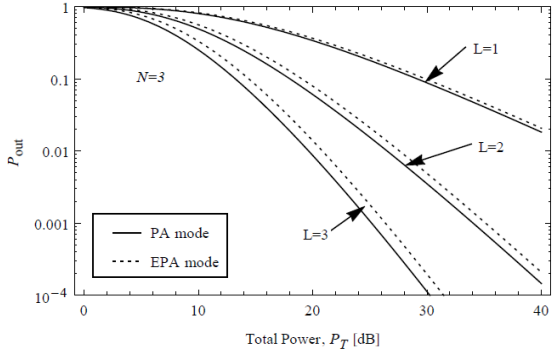

**Fig. 5.** Effect of the PA and EPA modes on the outage performance of multihop systems with diversity reception over $n*$Rayleigh fading channels.

By introducing Lagrange multipliers, the approximate power allocation for the $i$th hop can be expressed as

$$P_i \approx P_T \left[\sum_{k=1}^{N} \frac{n_i \Gamma(Lm_i, X_i) X_k^{Lm_k} e^{-X_k}}{n_k \Gamma(Lm_k, X_k) X_i^{Lm_i} e^{-X_i}}\right]^{-1}. \quad (17)$$

where $X_i = \beta_i \gamma_{th}^{\frac{1}{n_i}}$. From (17), we obtain the power allocation for the $l$th diversity branch as $P_l = P_i/L$. For $N = 2$, (17) is simplified to

$$P_1 \approx \frac{P_T}{\left[1 + \frac{n_1 \Gamma(Lm_1, X_1) X_2^{Lm_2} e^{-X_2}}{n_2 \Gamma(Lm_2, X_2) X_1^{Lm_1} e^{-X_1}}\right]}. \quad (18)$$

A similar equation can be written in terms of $P_2$. We note that (17) is a transcendental function and it is challenging to derive a closed-form for the transmitted power per hop. Hence, we calculate it numerically using a root-finding algorithm such as the bisection, Newton or successive numeric approximation methods.

*Asymptotic Analysis:* To gain further insight into the performance of regenerative systems over $n*$Rayleigh fading channels, we compute an asymptotic solution for (17) using the fact that $x^\alpha e^{-x} = \Gamma(\alpha + 1, x) - \alpha \Gamma(\alpha, x)$ [14, eq.(8.356.2)] and $\Gamma(\alpha + 1, x) \leq \alpha \Gamma(\alpha)$ for small values of $x$, to obtain

$$P_i \approx \frac{m_i}{n_i \sum_{k=1}^{N} \frac{m_k}{n_k}} P_T. \quad (19)$$

From (19), it can be seen that the power allocation for the $i$th hop depends only on the fading severity parameters, regardless of the channel statistics ($\lambda_i$). This means that when $P_T$ is set high, the power allocation for the $i$th hop is high, consequently, the power allocated for each diversity branch increases and is reduced as $L$ increases. In this case, the shadowing effects or path loss (related to $\lambda_i$) are negligible, corresponding to the same situation when the distances between successive nodes are approximately equal, resulting in a similar path loss on all the nodes.

Fig. 5 depicts the minimum outage probability versus the total power consumed to transmit the source's message to the destination node. In this scenario, we assume a three-hop system with/without diversity reception and channel quality with $\lambda_1 = 1, \lambda_2 = 1$ and $\lambda_3 = 10$ for $n_1 = 3, n_2 = 3$ and $n_3 = 2$, respectively. It is clear that the power allocation is more beneficial if diversity reception is used. For instance, at $L = 3$, the power allocation ratio $\rho = P_i/P_T$ is evaluated from (17) using the successive approximation algorithm, which turns out to be $\rho \approx 0.44, 0.44$, and $0.14$ for $n_1, n_2$, and $n_3$, respectively. It should be noted here that the regenerative systems allocate larger power to the weaker links to reduce the overall outage probability compared to the equal power allocation (EPA) mode where the total transmitted power splits equally between the nodes ($P_i = P_T/3$). On the other hand, the PA ratio for the case of no-diversity reception, is calculated as $\rho \approx 0.41, 0.41,$ and $0.184$ for $n_1, n_2$, and $n_3$, respectively. We note that the underlying scheme allocates less power to the first two hops, approaching the EPA mode. As a result, the performance improvement is negligible for the no-diversity reception case.

## IV. CONCLUSION

In this letter, we analyzed the end-to-end performance of multihop-IVC systems with regenerative and nonregenerative relays. In particular, we investigated the performance of both regenerative and nonregenerative systems with diversity reception over $n*$Rayleigh fading channels. We derived new closed-form expressions for the outage probability and the amount of fading. The power optimization problem has also been formulated and solved. Numerical results have shown that at high cascading order $n$, nonregenrative systems achieve outage performance close to that of regenerative systems. Finally, we demonstrated that optimizing the transmit power allocation for diversity reception systems can provide a significant performance gain compared to the equal power allocation scenario.

## REFERENCES

[1] M. O. Hasna and M. S. Alouini, "Outage probability of multihop transmission over Nakagami fading channels," *IEEE Commun. Lett.*, vol. 7, no. 5, pp. 216–218, May 2003.
[2] G. K. Karagiannidis, T. A. Tsiftsis, and R. K. Mallik, "Bounds of multihop relayed communications in Nakagami-$m$ fading," *IEEE Trans. Commun.*, vol. 54, no. 1, pp. 18–22, Jan. 2006.
[3] I. Z. Kovács, "Radio Channel Characterization for Private Mobile Radio Systems," Ph.D. dissertation, Aalborg University, Center for Person-Kommunikation, 2002.
[4] J. Salo, "Statistical analysis of the wireless propagation channel and its mutual information", Doctoral thesis, Helsinki University of Technology, Espoo, Finland, Jul.2006.
[5] D. Chizhik, G. Foschini, M. Gans, and R. Valenzuela, "Keyholes, correlations and capacities of multielement transmit and receive antennas," *IEEE Trans. Wireless Commun.*, vol. 1, pp. 361–368, Apr. 2002.
[6] D. Gesbert, H. Bölcskei, D. A. Gore, and A. J. Paulraj, "Outdoor MIMO wireless channels: Models and performance prediction," *IEEE Trans. Commun.*, vol. 50, no. 12, pp. 1926–1934, Dec. 2002.
[7] D. W. Matolak and J. Frolik, "Worse-than-Rayleigh fading: Experimental results and theoretical models," *IEEE Commun. Mag.*, vol. 49, no. 4, pp. 140-146, Apr. 2011.
[8] A. F. Molisch, F. Tufvesson, J. Karedal, and C. F. Mecklenbräuker, "A survey on vehicle-to-vehicle propagation channels," IEEE Wireless Commun., vol. 16, no. 6, pp. 12–22, Dec. 2009.
[9] D. W. Matolak, Q. Wu, and I. Sen, "5 GHz band vehicle-to-vehicle channels: Models for multiple values of channel bandwidth," IEEE Trans. Veh. Technol., vol. 59, no. 5, pp. 2620–2625, Jun. 2010.
[10] Seyfi, M., Muhaidat, S, Jie Liang, and Uysal, M., "Relay Selection in Dual- Hop Vehicular Networks," *IEEE Signal Processing Letters*, vol.18, no.2, pp.134,137, Feb. 2011.
[11] H, Yun, *et al*. "Partial relay selection for a roadside-based two-way amplify-and-forward relaying system in mixed Nakagami-m and 'double' Nakagami-m fading." *IET Communications*, vol.8, no.5 pp571-577, Mar 2014.
[12] H. Lu, Y. Chen, and N. Cao, "Accurate approximation to the PDF of the product of independent Rayleigh random variables," *IEEE Antennas Wireless Propag. Lett.*, vol. 10, pp. 1019–1022, Oct. 2011.
[13] M. K. Simon and M.-S. Alouini, *Digital Communication Over Fading Channels*, 2nd ed. New York: Wiley, 2004.
[14] I. S. Gradshteyn and I. M. Ryzhik, *Table of Integrals, Series and Products*, 7th ed. New York: Elsevier, 2007.
[15] J. Salo, H. E. Sallabi, and P. Vainikainen, "The distribution of the product of independent Rayleigh random variables," *IEEE Trans. Antennas Propagat.*, vol. 54, no. 2, pp. 639–643, Feb. 2006.
[16] E. K. Al-Hussaini and A. A. M. Al-Bassiouni, "Performance of MRC diversity systems for the detection of signals with Nakagami fading," *IEEE Trans. Commun.*, vol. COM-33, pp. 1315–1319, Dec. 1985.
[17] The wolfram functions site. [Online] http://functions.wolfram.com/